# Monitoring of saline tracer movement with vertically distributed self-potential measurements at the HOBE agricultural test site, Voulund, Denmark

D. Jougnot[1,2], N. Linde[1], E. B. Haarder[3], M. C. Looms[3]

1. Applied and Environmental Geophysics Group, Institute of Earth Sciences, University of Lausanne, 1015 Lausanne, Switzerland.
2. Now at CNRS, UMR 7619, METIS, F-75005 Paris, France
3. Department of Geosciences and Natural Resource Management, University of Copenhagen, DK-1350 Copenhagen, Denmark.

**Authors contact:**

Damien Jougnot: damien.jougnot@unil.ch

Niklas Linde: niklas.linde@unil.ch

Eline B. Haarder: eh@geo.ku.dk

Majken C. Looms: mcl@geo.ku.dk

**Abstract**

The self-potential (SP) method is sensitive to water fluxes in saturated and partially saturated porous media, such as those associated with rainwater infiltration and groundwater recharge. We present a field-based study at the Voulund agricultural test site, Denmark, that is, to the best of our knowledge, the first to focus on the vertical self-potential distribution prior to and during a saline tracer test. A coupled hydrogeophysical modeling framework is used to simulate the SP response to precipitation and saline tracer infiltration. A layered hydrological model is first obtained by inverting dielectric and matric potential data. The resulting model that compares favorably with electrical resistance tomography models is subsequently used to predict the SP response. The electrokinetic contribution (caused by water fluxes in a charged porous soil) is modeled by an effective excess charge approach that considers both water saturation and pore water salinity. Our results suggest that the effective excess charge evolution prior to the tracer injection is better described by a recent flux-averaged model based on soil water retention functions than by a previously proposed volume-averaging model. This is the first time that raw (i.e., without post-processing or data-correction) vertically distributed SP measurements have been explained by a physically based model. The electrokinetic contribution cannot alone reproduce the experimental SP data during the tracer test and an electro-diffusive contribution (caused by concentration gradients) is needed. The predicted amplitude of this contribution is too small to perfectly explain the data, but the shape is in accordance with the field data. This discrepancy is attributed to imperfect descriptions of electro-diffusive phenomena in partially saturated soils, unaccounted soil heterogeneity, and discrepancies between the measured and predicted electrical conductivities in the tracer infiltration area. This study opens the way for detailed long-term field-based investigations of the SP method in vadose zone hydrology.

# 1. Introduction

Quantification of water fluxes in the vadose zone is essential for many hydrological and environmental applications. Classical approaches based on matric potentials or tracer test data (e.g., Vereecken et al., 2008; Tarantino et al., 2009) are limited by the punctual nature of such measurements. The data might be strongly influenced by local heterogeneities and water fluxes are obtained indirectly by differencing. Lunati et al. (2012) demonstrated that large errors occur when mass and energy balances are computed from discrete measurements. One way to overcome the influence of local heterogeneity is to use geophysical measurements that are representative of larger volumes (e.g., Hubbard and Linde, 2011). The present contribution focuses on the self-potential (SP) method and to what extent it can be used to infer water fluxes and tracer transport at an experimental research site. The self-potential method is non-invasive and sensitive to subsurface flow and transport processes (for reviews, see Jouniaux et al., 2009; Revil et al., 2012; Revil and Jardani, 2013). It is a passive method, in which spatial and temporal variations of the electrical potential field are measured with respect to a reference electrode. The recorded self-potential data are given as a superposition of several contributions, which makes interpretation challenging.

The electrokinetic (EK) contribution (often referred to as the streaming potential) is directly related to the water flux and the properties of the electrical double layer found at the mineral-pore water interface. Water flowing through the pore drags a fraction of the excess charge, which gives rise to a streaming current and a resulting electrical potential field. Electrokinetic effects have been studied for more than a century (Helmholtz, 1879) and are well understood in water saturated porous media (e.g., Jouniaux et al., 2009; Revil and Jardani, 2013). Two main approaches have been proposed to simulate streaming current generation at partial saturation. The first focuses on how the streaming potential coupling coefficient varies as a function of water saturation (e.g., Guichet et al., 2003; Revil and Cerepi, 2004; Darnet and Marquis, 2004; Allègre et al., 2010), while the second focuses on how the excess charge dragged by the water varies with water saturation (e.g., Linde et al., 2007a; Revil et al., 2007; Linde, 2009; Jackson, 2010; Mboh et al., 2012; Jougnot et al., 2012). The lack of agreement between researchers is partly due to a limited number of well-controlled experiments and the multiple contributions to the measured signal, including electrode effects (e.g., Jougnot and Linde, 2013).

Most well-controlled laboratory studies have focused on drainage experiments (Linde et al., 2007a; Allègre et al., 2010; Mboh et al., 2012) or drainage-imbition cycles (Haas and Revil, 2009; Vinogradov and Jackson, 2011; Jougnot and Linde, 2013; Allègre et al., 2014). Due to experimental difficulties, only few data sets describe the streaming potential coupling coefficient at partially saturated conditions at steady state (Guichet et al., 2003; Revil and Cerepi, 2004; Revil et al., 2007). These measurements suggest that the dependency on the streaming potential coupling coefficient with saturation is media dependent. Doussan et al. (2002) instrumented a lysimeter and monitored SP data and vertical water flux in partially saturated conditions. A limited number of hydrological studies have used SP monitoring data conducted at the field scale (e.g., Thony et al., 1997; Perrier and Morat, 2000; Revil et al., 2002; Rizzo et al., 2004; Suski et al., 2006; Maineult et al., 2008; Linde et al., 2011) and none of these were instrumented to measure the vertical distribution of the SP signal.

The electro-chemical contribution to the self-potential signal can have two different origins: oxido-reductive (redox) and electro-diffusion (diff) processes. Even if strong redox potential contrasts exist in the near surface, redox phenomena only contribute to the SP signals when an electronic conductor exists that connects two regions with different oxidation potential (e.g., metallic bodies, certain kinds of bacteria). If this is the case, the redox contribution is typically the larger contribution to the SP signal (e.g., Naudet et al., 2004; Linde and Revil, 2007). If not, the corresponding contribution is null (e.g., Hubbard et al., 2011). The electro-diffusive contribution occurs in the presence of concentration gradients and is linked to the differential diffusion of ions with different mobilities. Electro-diffusive phenomena have been extensively studied in saturated porous media (e.g., Maineult et al., 2004, 2005, 2006; Revil et al., 2005; Straface and De Biase, 2013), but only few works concern partially saturated conditions (Revil and Jougnot, 2008; Jougnot and Linde, 2013). In the past, the electro-diffusive contribution has often been ignored, for example, during SP monitoring of saline tracer tests (e.g., Bolève et al., 2011).

We present the first results of a long-term monitoring program designed to investigate the role of SP data for predictive *in situ* estimation of vertical water flux. The HOBE agricultural test site in Voulund (Denmark) was chosen as the vadose zone is extensive, flow and transport processes can be assumed to be mainly vertical, and it is well instrumented with meteorological, hydrological and geophysical tools and sensors (Jensen and Illangasekare, 2011). Vertically distributed non-polarizable electrodes that were installed at the site were monitored for more than 2 years. We first obtain a numerical model of the test site that is used to simulate water fluxes, ionic transport and SP signals. We then compare the predictions for different competing models of SP signal generation and place particular focus on the signal contributions (i.e., electrokinetic and electro-diffusive) during a saline tracer test.

## 2. Theoretical framework

Below, we present the theory used to describe water flow, transport and SP signal generation under the assumption of vertical flow and transport only.

### 2.1. Flow and transport

The vadose zone is the region comprised between the land surface and the water table. The water saturation, $S_w$, is defined as the ratio between the water and pore volumes: $S_w = \theta_w / \phi$, where $\theta_w$ is the volumetric water content (m$^3$ m$^{-3}$) and $\phi$ the medium porosity (m$^3$ m$^{-3}$). The effective water saturation is defined as:

$$S_e = \frac{\theta_w - \theta_w^r}{\phi - \theta_w^r}, \tag{1}$$

where $\theta_w^r$ is the residual water content (m$^3$ m$^{-3}$). The water retention function relates the effective saturation of the medium to its matric potential, $h$ (m). In this work, we use the van Genuchten (1980) model:

$$S_e = \left[1 + \left(\alpha_{VG} h\right)^{n_{VG}}\right]^{-m_{VG}}, \tag{2}$$

where $\alpha_{VG}$ (m$^{-1}$) is proportional to the inverse of the air-entry pressure, while $n_{VG}$ and $m_{VG} = 1 - \left(1/n_{VG}\right)$ are curve shape parameters.

Water fluxes are described by Richards' equation and the van Genuchten-Mualem model (Van Genuchten, 1980) is used for the relative permeability function, $k_w^{rel}$,

$$k_w^{rel}(S_e) = \sqrt{S_e}\left[1 - \left(1 - S_e^{1/m_{VG}}\right)^{m_{VG}}\right]^2. \tag{3}$$

The hydraulic conductivity as a function of saturation is then given as,

$$K_w(S_e) = k_w^{rel}(S_e) K_w^{sat}, \tag{4}$$

with $K_w^{sat}$ (m s$^{-1}$) the saturated hydraulic conductivity. The vertical water flux $u$ (m s$^{-1}$) is described by,

$$u = -K_w(S_e)\left(\frac{\partial h}{\partial z} + 1\right). \tag{5}$$

The pore water is an electrolyte containing $N$ ionic species $j$ with a concentration $C_j$ (mol L$^{-1}$). Transport under partial saturation for each species is driven by the water flux through the 1D conservation equation:

$$\frac{\partial(\theta_w C_j)}{\partial t} + \frac{\partial}{\partial z}\left[-\left(\alpha_z \frac{u}{\theta} + D_j^{eff}\right)\frac{\partial C_j}{\partial z} + u C_j\right] = 0, \tag{6}$$

where $\alpha_z$ (m) is the dispersivity of the medium along the $z$-axis, and $D_j^{eff}$ (m$^2$ s$^{-1}$) is the ionic diffusion coefficient of the $j^{th}$ ionic species in the porous medium.

## 2.2. Self-potential generation

The SP response of a given source current density $\mathbf{J}_S$ (A m$^{-2}$) can be described by two equations (Sill, 1983),

$$\mathbf{J} = \sigma \mathbf{E} + \mathbf{J}_S, \tag{7}$$

$$\nabla \cdot \mathbf{J} = 0, \tag{8}$$

where $\mathbf{J}$ (A m$^{-2}$) is the total current density, $\sigma$ (S m$^{-1}$) the bulk electrical conductivity of the medium, $\mathbf{E} = -\nabla\varphi$ (V m$^{-1}$) the electrical field, and $\varphi$ (V) the electrical potential. In absence of external source currents (i.e. no current injection in the medium), Eqs. (7) and (8) can be combined to obtain:

$$\nabla \cdot (\sigma \nabla \varphi) = \nabla \cdot \mathbf{J}_S. \tag{9}$$

The measured SP response at the $i^{th}$ electrode is the potential difference with respect to the reference electrode: $SP = \varphi_i - \varphi_{ref}$. In this work, we only consider electrokinetic (superscript EK) and electro-diffusive (superscript diff) contributions. Considering only 1D vertical variations, these sources can be summed to obtain the total source current density: $J_S = J_S^{EK} + J_S^{diff}$.

The electrokinetic source ($J_S^{EK}$) is directly related to the water flux. This source current density is traditionally defined with respect to the hydraulic gradient through the streaming potential coupling coefficient (Helmholtz, 1879), $C_{EK}$ (V m$^{-1}$):

$$J_S^{EK} = \sigma C_{EK} \frac{\partial h}{\partial z}. \tag{10}$$

No model has so far been able to predict its dependence on water saturation, $C_{EK}(S_w)$, for all published data (e.g., Guichet et al., 2003; Revil and Cerepi, 2004; Allègre et al., 2010; Vinogradov and Jackson, 2011).

An alternative approach to $C_{EK}$ proposed by Revil and Leroy (2004) for water saturated conditions was later extended to partial saturation by Linde et al. (2007a). It is based on the excess charge density, $\bar{Q}_v$ (C m$^{-3}$), in the pore water that helps to counterbalance the electrical charges at mineral surfaces (i.e. the charge fraction located in the so-called Gouy-Chapman diffuse layer, Fig. 1a). This excess charge distribution follows a Boltzmann distribution with properties that depend mainly on the pore water salinity and the electrical potential at the mineral surface (Hunter, 1981). When the salinity increases, the thickness of the diffuse layer shrinks (Fig. 1b). When the water flows in the pores (Fig. 1c), the excess charge is dragged in the medium and generates a corresponding current source density. From a variable change in Eq. (10), the electrokinetic source current density can be defined by:

$$J_S^{EK} = \bar{Q}_v^{eff} u, \qquad (11)$$

where $\bar{Q}_v^{eff}$ is the excess charge (in C m$^{-3}$) that is effectively dragged in the medium by the water flux $u$. This concept offers also an alternative definition of the streaming potential coupling coefficient (Revil and Leroy, 2004):

$$C_{EK} = -\frac{\bar{Q}_v^{eff}}{\sigma} \frac{K_w}{\rho_w g}, \qquad (12)$$

where $\rho_w$ (kg m$^3$) is the water density and $g$ the gravity acceleration (9.81 m s$^{-2}$). Titov et al. (2002) presented laboratory data that strongly suggest an inverse relationship between the effective excess charge, $\bar{Q}_v^{eff,sat}$, and permeability, $k$ (m$^2$). Jardani et al. (2007) proposed the following empirical expression for this type of relationship

$$\log_{10}\left(\bar{Q}_v^{eff,sat}\right) = -0.82 \log_{10}(k) - 9.23. \qquad (13)$$

This relationship has been successfully applied in several studies (e.g., Bolève et al., 2009; Revil and Mahardika, 2013), even if the influence of the pore water chemistry on $\bar{Q}_v^{eff,sat}$ is neglected. The effective excess charge depends on the salt concentration in the water phase $C_w$ (mol m$^{-3}$) as indicated in Fig. 1b (e.g., Hunter, 1981; Revil et al., 1999). This relationship has been observed experimentally for measurements of streaming potential coupling coefficients at different pore water conductivity, $\sigma_w$ (S m$^{-1}$) (i.e., salinity), by Pengra et al. (1999, their Fig. 9). Linde et al. (2007b) propose an empirical relationship to correct for the salinity effect at saturation that is based on literature data for various media:

$$\log_{10}\left(C_{EK}^{sat}\right) = a + b \log_{10}(\sigma_w) + c \left[\log_{10}(\sigma_w)\right]^2 \qquad (14)$$

where $a$ = -0.895, $b$ = -1.319, and $c$ = -0.1227 are best fit parameters.

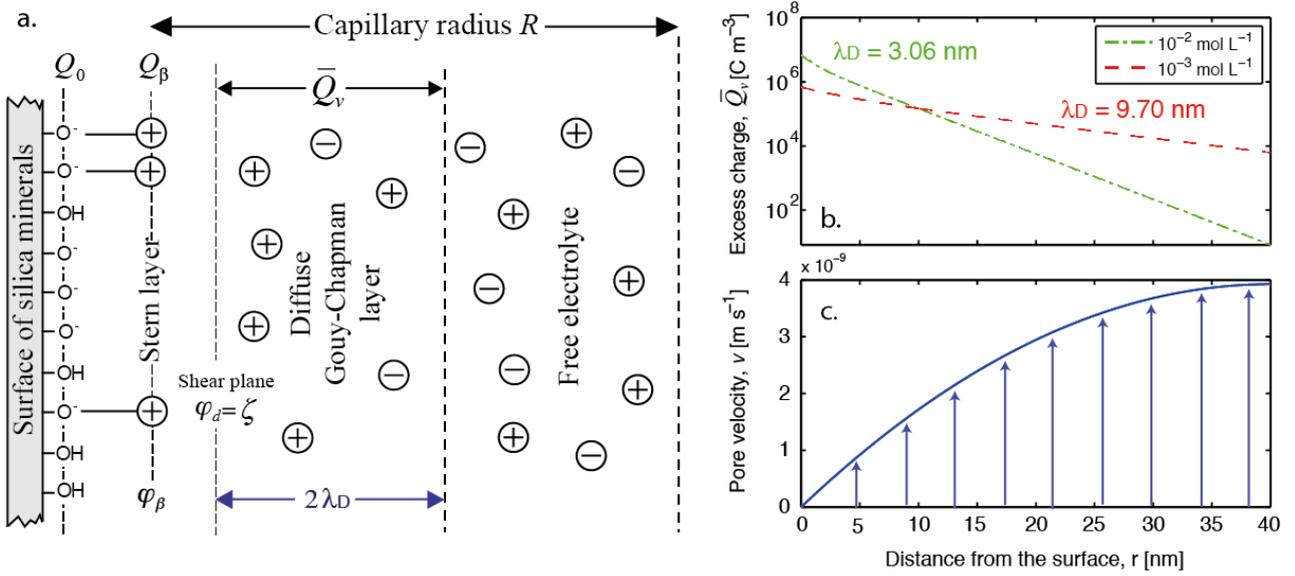

**Figure 1:** Effect of the NaCl electrolyte concentration upon the distribution of excess charge in a capillary with a 40 nm radius: (a) sketch of the electrical double layer, (b) excess charge and (c) pore water velocity distribution. A pore water salinity change from $C_w = 10^{-2}$ to $10^{-3}$ mol L$^{-1}$ result in the Debye lengths increasing from $\lambda_D = 3.06$ to $9.70$ nm. The electrokinetic coupling takes place in the diffuse Gouy-Chapman layer, which has a thickness on the order of two Debye lengths.

Based on a volume averaging approach, Linde et al. (2007a) consider that as the volume of water diminishes in the medium (i.e., the saturation decreases), and given that the amount of surface charge stays constant, the excess charge density $\bar{Q}_v^{\text{eff}}(S_w)$ should increase in the pore water. Linde et al. (2007a) propose the following scaling relationship:

$$\bar{Q}_v^{\text{eff}}(S_w) = \frac{\bar{Q}_v^{\text{eff,sat}}}{S_w}. \qquad (15)$$

This relationship provides satisfactory predictions when used to interpret laboratory SP data from drainage and imbibition experiments in rather homogeneous media (e.g., Linde et al., 2007a; Mboh et al., 2012; Jougnot and Linde, 2013). Nevertheless, this model is limited when applied to heterogeneous media, especially at low saturations. Indeed, the volume-averaging is only strictly valid when all pores have the same size (see Linde et al., 2011; Jougnot et al., 2012).

Jougnot et al. (2012) developed two flux averaging approaches that conceptualize the porous medium as a bundle of capillaries (see also, Linde, 2009; Jackson, 2010; Jackson and Leinov, 2012). Jougnot et al. (2012) upscale the electrokinetic properties of a given capillary with radius $R$ that contains a pore water with a given solute concentration, $C_w$, (i.e., the effective excess charge in the capillary $\bar{Q}_v^{\text{eff,R}}(R, C_w)$) to a representative elementary volume of a porous medium characterized by a saturated capillary size distribution that varies with saturation. This distribution $f_D$ can either be derived from the water retention function (referred to as the WR approach in the following) or from the relative permeability function (referred to as the RP approach in the following). The effective excess charge is obtained by integrating the distribution of the pore water flux $v^R(R)$ (m s$^{-1}$) and the distribution of excess charge density $\bar{Q}_v^{\text{eff,R}}(S_w, C_w)$ (C m$^{-3}$) within the capillaries:

$$\bar{Q}_v^{\text{eff}}(S_w, C_w) = \frac{\int_{R_{\min}}^{R_{S_w}} \bar{Q}_v^{\text{eff,R}}(R, C_w) v^R(R) f_D(R) dR}{\int_{R_{\min}}^{R_{S_w}} v^R(R) f_D(R) dR}. \qquad (16)$$

Flux averaging based on the WR or the RP approach ($f_D^{\text{WR}}$ or $f_D^{\text{RP}}$) yield two different effective excess charge functions ($\bar{Q}_v^{\text{eff,WR}}$ or $\bar{Q}_v^{\text{eff,RP}}$). Previous studies (Jougnot et al., 2012; Jougnot and Linde, 2013) indicate that the RP approach reproduce the observed amplitudes well, but is less accurate in describing the relative changes with saturation. The WR approach tends to underestimate the observed amplitudes, but reproduces relative changes with saturation rather well. In analogy with the relative permeability function (i.e., Eq. (4)), we decompose the effective charge function into a value at saturation, $\bar{Q}_v^{\text{eff,sat}}$, and a relative saturation-dependent function, $\bar{Q}_v^{\text{eff,rel}}(S_w)$:

$$\bar{Q}_v^{\text{eff}}(S_w) = \bar{Q}_v^{\text{eff,rel}}(S_w) \bar{Q}_v^{\text{eff,sat}}. \qquad (17)$$

The value of $\bar{Q}_v^{\text{eff,sat}}$ can be obtained by measurements, by using an empirical law (e.g. Eqs. (13) or (14)), or by inversion. The $\bar{Q}_v^{\text{eff,rel}}(S_w)$ is here derived using either the RP or the WR approach (see Eq. (16)).

The electro-diffusive source current density, $J_S^{\text{diff}}$, is generated by ionic charge separation between anions and cations with different mobilities $\beta_j$. It can be described by the microscopic Hittorf number which represents the fraction of the total current transported by a given ionic species, $j$:

$$t_j^H = \frac{\beta_j}{\sum_{\iota=1}^{Q} \beta_\iota}. \qquad (18)$$

Charged porous media can act as semi-permeable membranes that enhance or decrease this effect and this necessitates a macroscopic Hittorf number $T_j^H$. Revil and Jougnot (2008) describe how $T_j^H$ varies with water saturation and proposed an application to clay-rock. The electro-diffusive source current density is given by,

$$J_S^{\text{diff}} = -k_B T \sum_{j=1}^{Q} \left( \frac{T_j^H(S_w)}{q_j} \sigma(S_w) \right) \frac{1}{C_j} \frac{\partial C_j}{\partial z}, \qquad (19)$$

where $k_B = 1.38065 \times 10^{-23}$ J K$^{-1}$ is the Boltzman constant, $T$ (K) is the temperature, $q_j = \pm Z_j e_0$ (C) is the electrical charge of the considered ions, with $Z_j$ its valence and $e_0 = 1.3806 \times 10^{-19}$ C the elementary charge. The resulting electrical potential is referred to as electro-diffusive or junction potential (e.g., Maineult et al., 2005, 2006; Jouniaux et al., 2009), or membrane potential if the effect of the electrical double layer cannot be neglected (e.g., Revil et al., 2005). Jougnot and Linde (2013) showed that a saturation-independent microscopic Hittorf number $t_j^H$ (i.e., $T_j^H(S_w) = t_j^H$ in Eq. (19)) could explain experimental laboratory data.

From Eqs. (9), (11), and (19), the electrical problem can be simplified and re-written as:

$$\frac{\partial \varphi}{\partial z} = \frac{1}{\sigma(S_w)}\left[\bar{Q}_v^{\text{eff}}(S_w)u - k_B T \sum_{j=1}^{Q}\left(\frac{T_j^H(S_w)}{q_j}\sigma(S_w)\right)\frac{1}{C_j}\frac{\partial C_j}{\partial z}\right]. \quad (20)$$

The superposition principle allows decomposing the resulting SP signal into electrokinetic and electro-diffusive contributions $SP = SP^{\text{EK}} + SP^{\text{diff}}$ (e.g., Jougnot and Linde, 2013).

## 2.3. Electrical and dielectric petrophysical relationships

In hydrology, dielectric permittivity, $\varepsilon$ (F m$^{-1}$), and electrical conductivity, $\sigma$ (S m$^{-1}$), are often measured to infer soil water saturation and salinity (e.g., Huisman et al., 2003; Friedman, 2005; Vereecken et al., 2008; Laloy et al., 2011). Many petrophysical relationships exist to relate dielectric permittivity and water content (see Huisman et al., 2003 for a review). These relationships are often expressed in term of relative permittivity: $\varepsilon_r = \varepsilon/\varepsilon_0$ with $\varepsilon_0 = 8.854 \times 10^{-12}$ F m$^{-1}$ being the dielectric permittivity of vacuum. Among the existing relationships, one can distinguish between empirical relationships (e.g., Topp et al., 1980) and those based on mixing laws or up-scaling procedures (e.g., the Complex Refractive Index Model: Dobson et al., 1985; Roth et al., 1990). These more theoretical relationships are of interest as they can be used to explicitly account for the dielectric permittivities of the different components, the medium porosity, and its pore space geometry. In addition, temperature effects on the relative permittivity of pore water can be taken into account by the following empirical relationship (Weast et al., 1988):

$$\varepsilon_w(T[°C]) = 78.54\left[1 - 4.579\times 10^{-3}(T-25) + 1.19\times 10^{-5}(T-25)^2 - 2.8\times 10^{-8}(T-25)^3\right]. (21)$$

Linde et al. (2006) extended the volume averaging approach of Pride (1994) to describe the relative permittivity in partially saturated conditions:

$$\varepsilon_r = \phi^m \left[ S_w^{\ n} \varepsilon_w + \left(\phi^{-m} - 1\right)\varepsilon_s + \left(1 - S_w^{\ n}\right)\varepsilon_a \right], \tag{22}$$

where $\varepsilon_w$, $\varepsilon_s$, and $\varepsilon_a$ are the relative permittivities of water, grain and air, respectively. The petrophysical parameters *m* and *n* are the cementation and the saturation index defined by Archie (1942). They describe the pore space and the fluid phase geometry (i.e., tortuosity and constrictivity), respectively (Revil et al., 2007). Linde et al. (2006) also derived a petrophysical relationship to predict electrical conductivity:

$$\sigma = \phi^m \left[ S_w^{\ n} \sigma_w + \left(\phi^{-m} - 1\right)\sigma_s \right], \tag{23}$$

where $\sigma_w$ and $\sigma_s$ are the pore-water and the grain surface conductivities (S m$^{-1}$). This petrophysical relationship has been shown to reproduce variations of electrical conductivity with saturation (e.g., Breede et al., 2011; Laloy et al., 2011). When the grain surface conductivity can be neglected (e.g., typically in absence of clay), Eq. (23) can be simplified (Archie, 1942): $\sigma = \phi^m S_w^{\ n} \sigma_w$.

## 3. Material and methods

In this section, we describe the experimental set-up at the Voulund agricultural field site (Denmark), the strategy used for model calibration, and the numerical implementation of our modeling framework described in section 2.

### 3.1. Voulund test site at HOBE and set-up

The Danish hydrological observatory (HOBE) is focused on the Skjern River catchment (Jensen and Illangasekare, 2011) (Fig. 2a). In this catchment, the Voulund field site was chosen for detailed investigations of surface-aquifer processes in an agricultural environment. This area that is the focus of the present study is characterized by a relatively flat surface topography. The field site is equipped for meteorological, hydrogeological and geophysical monitoring (Fig. 2b and c). A full description of the observatory and the Voulund field site are available online (www.hobe.dk).

At Voulund, the geology is characterized as a fairly homogeneous sandy soil. The water table is monitored on site and it is located between 5.5 and 6.5 m depth. The soils comprising the vadose zone were characterized by five drilling campaigns between September 2011 and April 2012. 6 m long drill cores were extracted from the site and cut into 7.7 cm long samples. The sediment samples were weighed both after collection, and after 48 hours in an oven to determine the volumetric water content, total porosity and saturation. Additionally, 71 samples from one of the five cores were analyzed for grain size distribution. Over the 6 first meters, the soil is mainly composed by more than 90 % of very fine to very coarse sand. Although, between 5 and 10 % of silt and clay can be found in layer 1 and 2, and less than 5 % in layer 4 and 5. Details about the drilling campaigns and core sample analysis can be found in Uglebjerg (2013). Furthermore, laboratory measurements were performed on 100 cm³ soil cores extracted in the near vicinity of the field site. Retention characteristics at 6 depths down to 2.05 m and the soil hydraulic conductivities at 3 depths down to 0.8 m were determined (Vasquez, 2013). In general, the sediments constituting the soil can be divided into 7 different layers (Fig. 2c and Table 1).

**Table 1:** Gaussian prior for the soil properties of each layer in the MCMC inversion: the mean value is expressed on the first line, while the standard deviation is given in parenthesis.

| Layer | Depth | $\theta_w^r$ (-) | $\phi$ (-) | $\log_{10}(\alpha_{VG})$ (m$^{-1}$) [a] | $n_{VG}$ (-) | $\log_{10}(K_w^{sat})$ (m s$^{-1}$) [a] | $m$ (-) [b] | $n$ (-) [b] |
|---|---|---|---|---|---|---|---|---|
| 1 | 0 – 0.20 | 0.05 (0.01) | 0.39 (0.01) | -0.87 (0.50) | 1.36 (0.10) | 3.31 (1.47) | | |
| 2 | 0.20 – 0.45 | 0.04 (0.01) | 0.38 (0.02) | -1.15 (0.32) | 2.30 (0.36) | 2.74 (0.49) | | |
| 3 | 0.45 – 0.80 | 0.03 (0.01) | 0.38 (0.02) | -1.29 (0.10) | 2.30 (0.36) | 2.44 (0.46) | | |
| 4 | 0.80 – 1.20 | 0.07 (0.03) | 0.40 (0.02) | -1.02 (0.17) | 1.71 (0.31) | 2.13 (0.12) | 1.40 (0.10) | 2.00 (0.30) |
| 5 | 1.20 – 1.75 | 0.06 (0.02) | 0.37 (0.04) | -0.96 (0.14) | 2.49 (0.64) | 2.46 (0.43) | | |
| 6 | 1.75 – 2.55 | 0.04 (0.01) | 0.39 (0.02) | -1.21 (0.36) | 2.89 (1.04) | 2.46 (0.43) | | |
| 7 | 2.55 – 7.30 | 0.04 (0.01) | 0.39 (0.02) | -1.29 (0.22) | 2.89 (1.04) | 2.42 (0.43) | | |

a. Given the large parameter range, a log-normal distribution has been considered for these parameters
b. For the inversion using the volume averaging model (see Eq. (22); one value for all layers)

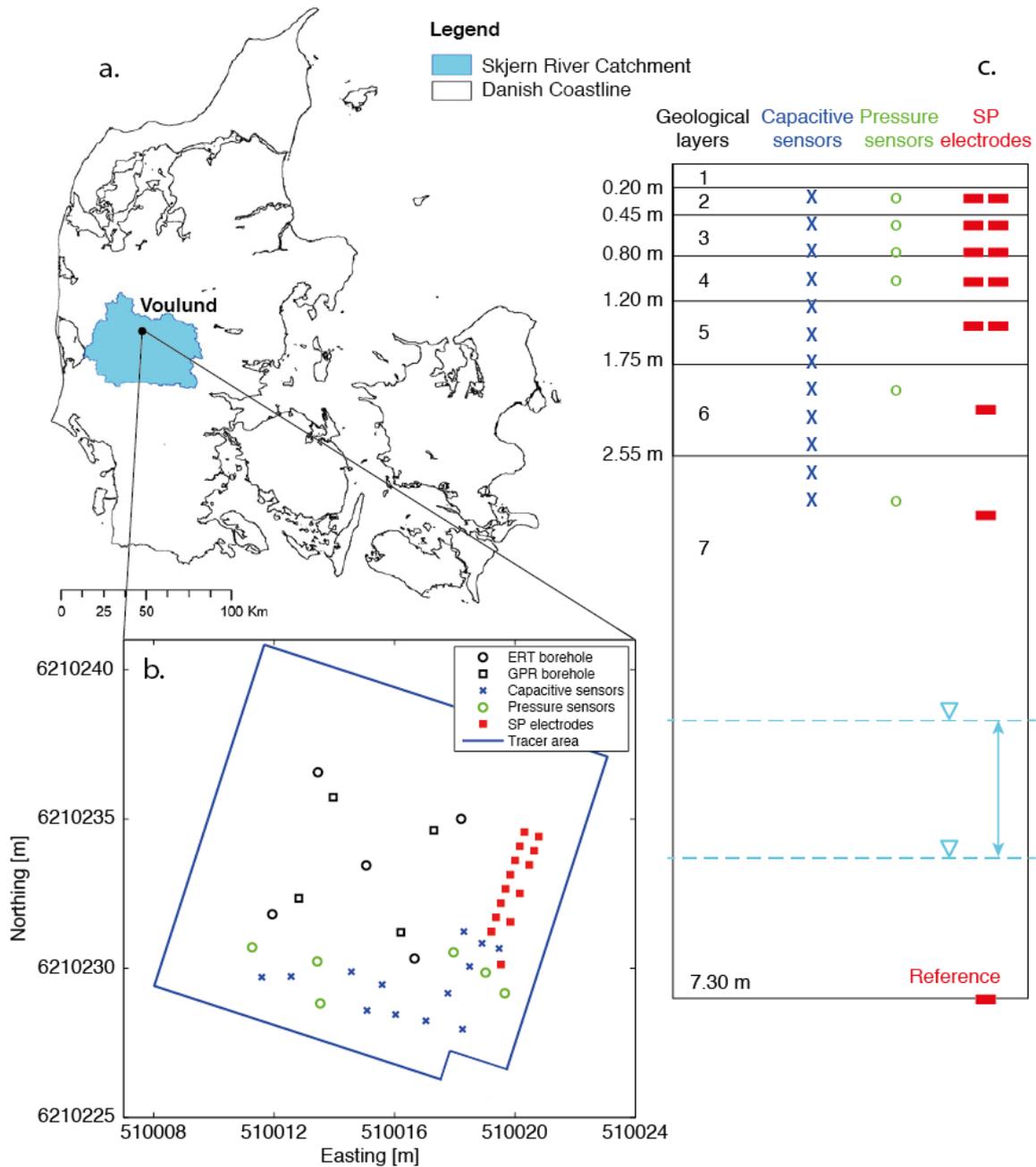

**Figure 2:** (a) Location of the Voulund test site in the Skjern river catchment, Denmark (modified from Jensen and Illangasekare, 2011). (b) Overview of sensor positioning and (c) sensor depths. The blue dashed lines in (c) correspond to the maximum and minimum observed water level variation during the study.

In the present study, meteorological and hydrological data were used as input to characterize the soil and its state evolution through a parameter inversion procedure. Due to the lateral homogeneity of the soil and the absence of significant topography, hydrological processes are expected to be primarily vertical. We used precipitations data from a ground level rain gauge and potential evapotranspiration calculated from solar radiation, wind speed, air temperature, and relative humidity using the Penman-Monteith equation assuming a 0.15 m vegetation cover (Monteith, 1965). These data were available as hourly values since 2009.

Monitoring was performed using 17 5TE and 6 MPS-1 Decagon sensors (www.decagon.com) down to 3 m depth. The 5TE sensors measure the relative dielectric permittivity ($\varepsilon_r$), the temperature ($T$ in °C) and the electrical conductivity ($\sigma$ in µS cm$^{-1}$) every 0.25 m from 0 to 3 m, while the MPS-1 sensors measure the matric potential ($h$ in m) at 0.25, 0.50, 0.75, 1.00, 2.00, and 3.00 m depth. These measurements were stored in a Campbell datalogger every 20 minutes since July 2011.

The field site was also instrumented to conduct cross-borehole geophysical monitoring with both electrical resistivity tomography (ERT) and ground penetrating radar (GPR) using a total of 9 boreholes of 6 m depth with PVC tubes installed in a square (Fig. 2b). In the current study, we consider the ERT measurements conducted in a five-borehole configuration. Four boreholes form a 5 m large square with an additional borehole at its center. 24 electrodes were installed in each borehole with a spacing of 0.25 m from 0.25 to 6.00 m depth. In addition, 12 electrodes were placed close to the surface (0.25 m depth) along the diagonals of the square. Each electrode consisted of a 5 cm wide stainless steel mesh, which was wrapped around a PVC tube. After lowering the PVC tubes into their correct location, the drill holes were backfilled with oven-dry sand from the bottom up while gently vibrating the tubes. This was done to avoid cavities along the PVC tubes and thereby ensuring a good electrical contact between the surrounding formation and the electrodes. ERT measurements were acquired on a daily to weekly basis for approximately one year. The measurement configuration details for injection and potential measurement electrodes are described by Haarder et al. (Accepted).

The 3D ERT data were inverted to obtain a 3D distribution of resistivity using the R3t inversion code (Binley, 2014). The area within the five boreholes was discretized using an unstructured tetrahedral mesh with a typical length scale of 0.125 m resulting in 614.698 elements. To obtain a 1D resistivity distribution as a function of depth the median and the first and third quartiles of the resistivity at a given depth was used in order to minimize the influence of inversion artefacts.

The SP monitoring was conducted using 15 non-polarizable Pb-PbCl$_2$ electrodes (Petiau, 2000) located at 0.25 to 3.20 m depth with the reference electrode at 7.30 m depth (Fig. 2c). Two electrodes were installed at every depth level in the shallow part (0.25 m, 0.50 m, 0.75 m, 1.00 m, and 1.45 m), while only one electrode was installed at every depth level in the deeper part (1.90 m, 2.50 m, 3.10 m, and 7.30 m). The deepest electrode was chosen as the reference electrode due to its position below the water table. The SP electrodes were installed on July 20$^{th}$ 2011 and the monitoring is running continuously since then with measurements stored every 5 minutes.

A saline tracer test was conducted in September 2011 to characterize water infiltration processes in situ and to test the ability of geophysical monitoring to quantify groundwater recharge (Haarder et al., Accepted). On the 14$^{th}$, the equivalent of 3 mm of brine ($\sigma_w^{Tr}$ = 230 mS cm$^{-1}$) was infiltrated within 2 hrs across a 142 m$^2$ area around the borehole square (Fig. 2b). This large tracer injection zone provides an injection that can be approximated as 1D in the vicinity of the geophysical sensors.

## 3.2. Hydrological model calibration

In a first step, we calibrate a hydrological model using the dielectric permittivity and matric potential measurements. Figure 3a describes how this was done using Markov chain Monte Carlo (MCMC) simulation. The prior distributions of the different parameters were chosen based on core sample analysis (Table 1). The hydrological parameters consist of porosities ($\phi$), saturated hydraulic conductivities ($K_w^{sat}$), residual water content ($\theta_w^r$), and van Genuchten parameters ($\alpha_{VG}$ and $n_{VG}$) that were assigned individually to each of the 7 layers. The petrophysical parameters depend on the relationship used to link dielectric permittivity ($\varepsilon_{sim}$) and water content ($\theta_w^{sim}$). Herein, we used the petrophysical relationship by Linde et al. (2006, see Eq. (22)). We included additional petrophysical parameters ($m$ and $n$) as priors in the MCMC procedure. It was assumed that these petrophysical parameters were the same for all soil layers and that surface conductivity could be neglected.

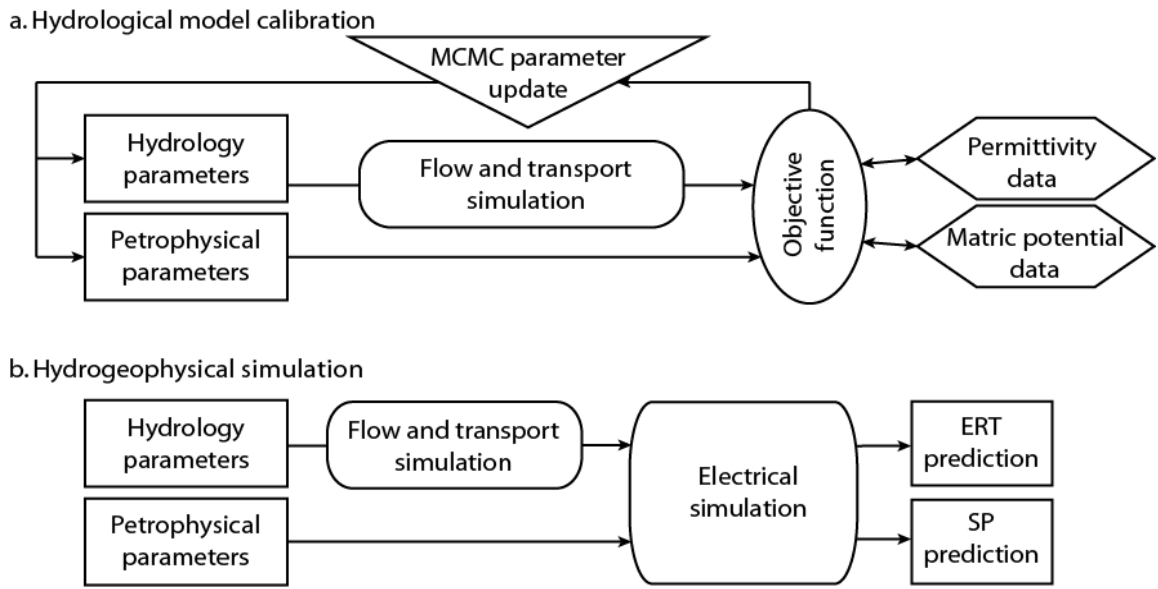

**Figure 3:** (a) Hydrological model calibration and (b) hydrogeophysical simulation schemes used in this work.

The hydrological forward problem (flow and transport) is solved using the Hydrus1D software (Simunek et al., 1998) with certain transport properties ($\alpha_z = 0.25$ m, and $D_{\text{NaCl}}^{\text{eff}} = 10^{-9}$ m$^2$ s$^{-1}$) assumed known (Haarder et al., Accepted). The 7 layers (see Fig. 2c) were discretized in 300 cells with appropriate boundary conditions assigned based on meteorological data (precipitation and potential evapotranspiration) and water table depth measured on site. In order to limit the influence of the initial conditions, we conducted simulations over 500 days starting on August 1$^{\text{st}}$ 2010; in the following, we consider the day that the tracer test was conducted to be Day 0. The simulation period is discretized in daily time steps (one meteorological input per day) with a refinement to hourly periods after Day -4 to better constrain the time evolution during the tracer test.

The MCMC inversion procedure uses the measured relative permittivity ($\varepsilon_{\text{meas}}$) and matric potential ($h_{\text{meas}}$) as observables. The DREAM$_{(ZS)}$ algorithm is used with three chains and standard values of algorithmic variables; see Laloy and Vrugt (2012) for details. In the inversion, we use a standard log-likelihood function that is proportional to:

$$SSR = \frac{1}{2}\sum\left(\frac{\varepsilon_{\text{meas}} - \varepsilon_{\text{sim}}}{\sigma_\varepsilon^{\text{std}}}\right)^2 + \frac{1}{2}\sum\left(\frac{h_{\text{meas}} - h_{\text{sim}}}{\sigma_h^{\text{std}}}\right)^2, \tag{24}$$

where the subscripts meas and sim indicate measured and simulated responses. Standard deviations of $\sigma_\varepsilon^{\text{std}} = 0.5$ and $\sigma_h^{\text{std}} = 0.25$ m were assigned for the relative permittivity and the matric potential. The rather large value for $\sigma_h^{\text{std}}$ is due to the rather poor quality of the MPS-1 data. This MCMC results represent a full posterior probability density function of the hydrodynamic parameters that explain the hydrologic measurements ($\varepsilon_{\text{meas}}$ and $h_{\text{meas}}$), but we only consider the best fit model in the following.

### 3.3. Hydrogeophysical simulation

The theoretical framework presented in section 2 was implemented to solve the coupled hydro-electrical problem numerically (Fig. 3b). The hydrologic forward model is solved using the best-fit hydrodynamic parameters (see section 3.2) to obtain the vertical distribution of the water flux ($u$ in m s$^{-1}$), the solute concentration ($C_w$ in mol L$^{-1}$), and the water saturation ($S_w$) as a function of time. The pore-water conductivity was computed from the solute concentration $\sigma_w(C_w)$ using the Sen and Goode (1992) relationship. It yields a soil electrical conductivity that depends on the water saturation and the solute concentration: $\sigma(S_w, C_w)$ based on Eq. (23).

The effective excess charge functions $\bar{Q}_v^{\text{eff}}(S_w, C_w)$ for each geological layer were calculated using the volume averaging model by Linde et al. (2007a), as well as the RP and WR approaches by Jougnot et al. (2012) using the hydrodynamic parameters obtained from the inversion procedure (see section 3.2). $\bar{Q}_v^{\text{eff}}$ is then calculated for each cell and time step based on the $S_w$ and $C_w$ distribution at the considered time. The electrokinetic contribution to the source current density $J_S^{\text{EK}}$ is given by Eq. (11). The electro-diffusive contribution, $J_S^{\text{diff}}$, is calculated from the solute concentration gradients and the medium's electrical conductivity (Eq. (19)).

Based on the distribution of electrical conductivity and the total source current density: $J_S = J_S^{\text{EK}} + J_S^{\text{diff}}$, the electrical problem is solved using a modified version of MaFloT (see maflot.com, Künze and Lunati, 2012). This yields the electrical potential vertical distribution, thus the SP signal from Eq. (20), at the different times (i.e., $SP(z,t)$).

## 4. Results and interpretations

### 4.1. Hydrological calibration results

The MCMC inversion procedure described in section 3.2 was applied for data covering a period of 70 days that included the tracer test injection and the monitoring (from Day -2 to Day +68). We initially considered three different petrophysical relationships in the inversion: the Topp equation, the CRIM, and the one proposed by Linde et al. (2006, Eq. (22)). The latter was retained for further analysis as it can describe both the dielectric permittivity (Eq. (22)) and the electrical conductivity (Eq. (23)) with the same medium parameters: $m$ and $n$. Figure 4 shows the best-fitting predictions of relative dielectric permittivity at different depths (the average RMSE is 0.55). Large $\varepsilon_{meas}$ peaks could not be reproduced as they would correspond to unphysical water contents (i.e., largely superior to porosity). This behavior is attributed to unaccounted salinity effects, similar to those studied by Rosenbaum et al. (2011). Unfortunately, we could not use their empirical correction models as our permittivities were much lower and the salinities were sometimes higher than the ranges considered by Rosenbaum et al. (2011). The average RMSE for matric potential is 0.37 m. This poor RMSE is likely due to a poor calibration of the sensors to the studied soil. The model parameters with the best fitting predictions are presented in Table 2. The hydrodynamic parameters are consistent with the laboratory characterization of core samples (Vasquez, 2013). The petrophysical parameters, $m = 1.38$ and $n = 1.57$, are also in good agreement with literature data for sand and sandy soils (e.g., Friedman, 2005; Linde et al., 2007a). In the absence of sensors below 3 m, we assume that these petrophysical parameters can also describe the deeper region (between 3 and 7.3 m).

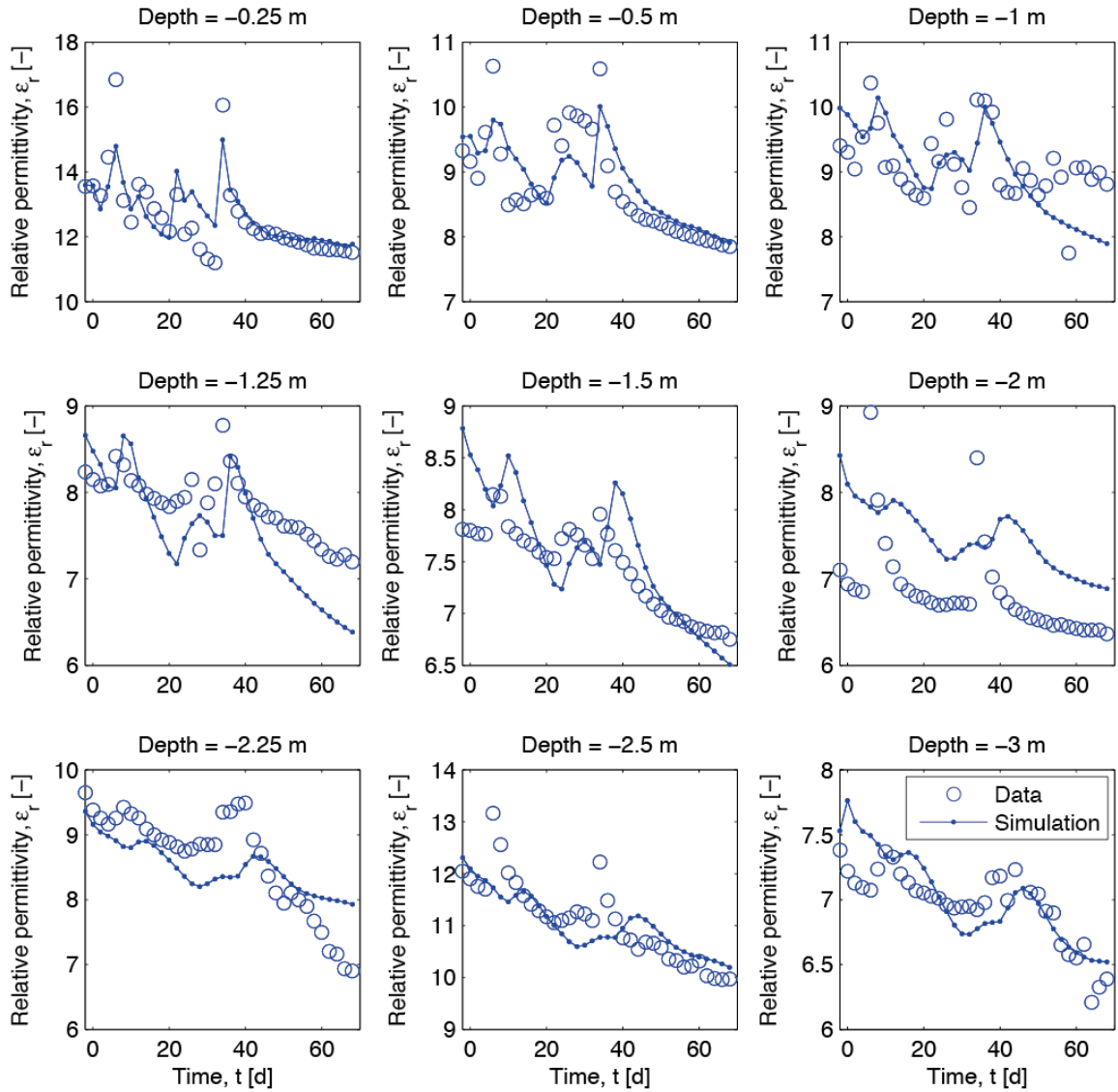

**Figure 4:** Comparison between the measured (empty circles) and simulated relative dielectric permittivity (filled dots) using the best fitting parameters from the hydrological inversion procedure. The temperatures measured by the 5TE ($T^{\text{meas}}$) are used to correct the simulated results using Eq. (21). The mean RMSE between the simulated and the measured relative permittivities is RMSE = 0.54.

**Table 2:** Best fit estimate of the soil parameters based on MCMC inversion using the Eq. (22) proposed by Linde et al. (2006).

| Layer | $\theta_w^r$ (-) | $\phi$ (-) | $\alpha_{VG}$ (cm$^{-1}$) | $n_{VG}$ (-) | $K_w^{sat}$ (cm d$^{-1}$) | $m$ (-) | $n$ (-) |
|---|---|---|---|---|---|---|---|
| 1 | 0.07 | 0.396 | 7.89 | 1.61 | 71 | | |
| 2 | 0.02 | 0.366 | 2.25 | 1.55 | 51 | | |
| 3 | 0.01 | 0.388 | 1.86 | 2.34 | 995 | | |
| 4 | 0.07 | 0.405 | 3.18 | 3.09 | 169 | 1.38 | 1.57 |
| 5 | 0.05 | 0.357 | 5.53 | 2.93 | 63 | | |
| 6 | 0.03 | 0.361 | 3.36 | 2.08 | 300 | | |
| 7 | 0.07 | 0.418 | 6.42 | 2.79 | 483 | | |

**4.2. Independent evaluation of the hydrological model**

Based on the estimated petrophysical parameters (section 4.1), we calculated the predicted electrical conductivity distribution as a function of depth ($\sigma(z)$) using Eq. (23) based on the simulated water saturation ($S_w$) and pore water salinity ($C_w$). We used the empirical expression by Sen and Goode (1992) to transform the water salinity into water conductivity ($\sigma_w$) while accounting for temperature effects. In the following, we will refer to electrical resistivity, the inverse of electrical conductivity: $\rho(z) = 1/\sigma(z)$. To evaluate the hydrological model and the petrophysical parameters obtained by the MCMC inversion, we compared our simulation results with selected electrical resistivity profiles obtain by inversion of 3D cross-borehole ERT measurements (Haarder et al., Accepted). To do so, we first simulate electrical resistivities based on our Hydrus 1D modeling results ($\rho(z)$). Corresponding electrical tomograms were calculated by forward simulation and subsequent inversion using the R2 code (Binley and Kemna, 2005). These results were then compared to the median, the first and the third quartiles of the models obtained from the field inversions. Prior to the tracer injection (Figure 5a), the general trend of the electrical resistivity distribution is well captured by our simulations. This independent evaluation provides some confidence in the hydrological model and the inferred petrophysical parameters. It also indicates that the simulated resistivity magnitudes are correct, which is crucial to accurately simulate SP magnitudes. After the tracer injection (Figure 5b-d), the hydrological simulations indicate a larger decrease of electrical resistivity in the tracer-affected region compared to the ERT results. It is well-known that ERT provides overly smooth images and it is difficult to assess to which extent the discrepancies in the upper 2 m are related to deficiencies in our hydrological model or to the limited resolution of the ERT results. A certain degree of care is needed when further interpreting results in this depth range.

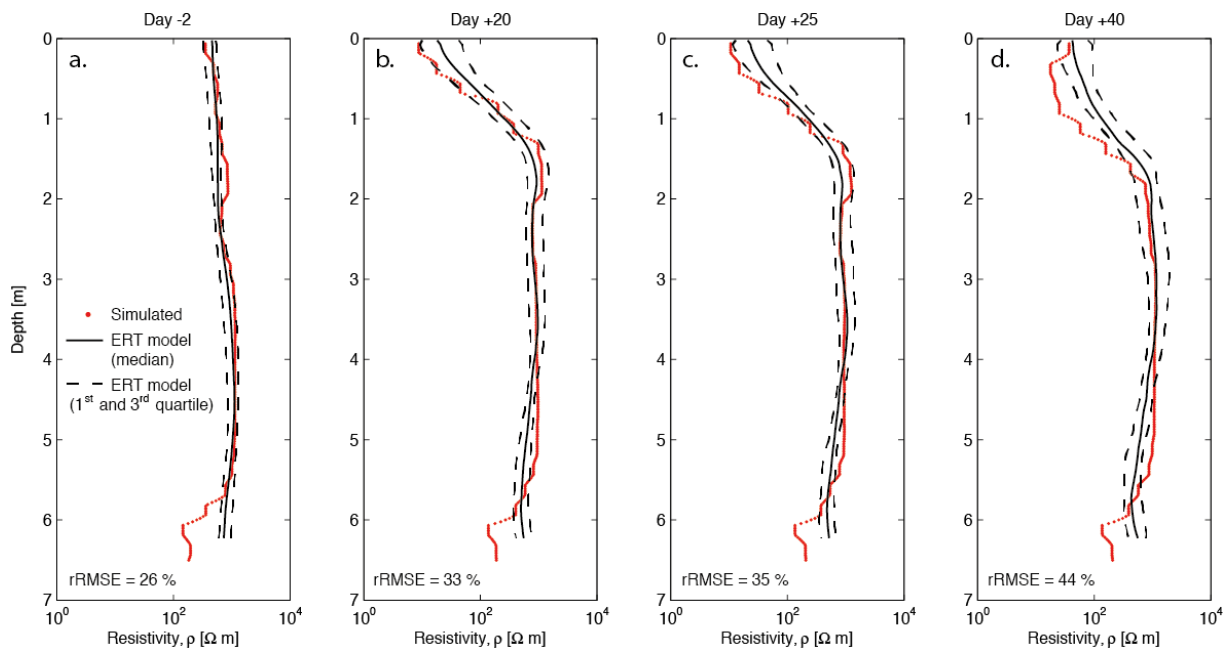

**Figure 5:** Comparison between inversion results describing the averaged resistivity models with depth. The red dots represent the results obtained from the best fitting hydrological simulation, while the black lines represent the field data inversions from cross-borehole ERT measurements: the solid line is the median of models while the dashed lines are the first and third quartiles of the models. The agreement is good, but the predicted models over-predict the reduction in resistivity due to the tracer injection. The relative RMSE (rRMSE) between the simulated and ERT model (normalized by the ERT model) are indicated for each day.

### 4.3. SP responses in natural conditions

We now turn our attention to the SP data and consider first the raw data (i.e., no filtering or detrending was applied) on Day -2 prior to the tracer injection (Figure 7). The amplitude ranges from $SP = -44.6$ mV at 0.25 m depth to -11.2 mV at 3.1 m depth. Similar shapes of the SP distribution have been predicted by numerical simulations (e.g., Linde et al., 2011; Jougnot et al., 2012), but never been measured in situ at more than two depths (see Doussan et al., 2002). The observed variations in the SP data at the same depth (maximum difference ~15 mV at 0.25 m depth) are most probably caused by local heterogeneities.

Prior to the tracer injection, the pore water concentration is assumed to be homogeneous ($\sigma_w = 200$ µS cm$^{-1}$ from the first drilling campaign) and the electro-diffusive contribution is null (i.e., $J_S^{diff} = 0$ A m$^{-2}$). In the following, we compare different models that describe $\bar{Q}_v^{eff} = f(S_w)$. Unfortunately, we did not have access to samples from the drilling campaign for measuring $\bar{Q}_v^{eff,sat}$. These samples were taken before the initiation of this work and they were destroyed during the grain size analyses. Instead, we relied on the empirical relationships proposed by Jardani et al. (2007) and Linde et al. (2007b) (Fig. 6a) to estimate $\bar{Q}_v^{eff,sat}$.

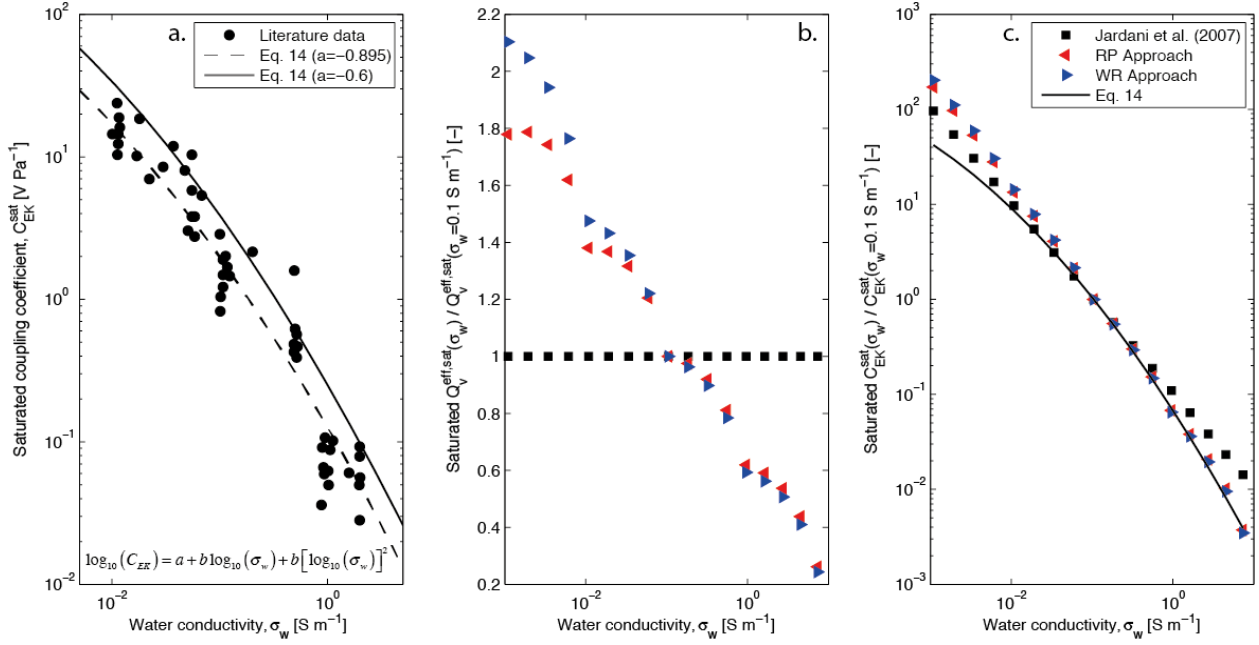

**Figure 6:** (a) Streaming potential coupling coefficient as a function pore water conductivity in saturated conditions (modified from Linde et al., 2007b). (b) Effective excess charge and (c) streaming potential coupling coefficient as a function of pore water conductivity $\sigma_w$ in saturated conditions relatively to their respective value for $\sigma_w = 0.1$ S m$^{-1}$ calculated using the different models in layer 4 and comparison to the models displayed in (a).

The model of Linde et al. (2007a) predicts that the effective excess charge increases with the inverse of saturation (Eq. (15)). The saturated effective excess charge density was calculated using the empirical relationship by Jardani et al. (2007; Eq. (13)) by using the layer permeabilities inferred by the inversion (Table 2). These estimates were subsequently divided by the water saturation at each depth to obtain $\bar{Q}_v^{\text{eff}}(z)$. Then, we solved the electrical problem for $SP^{\text{EK}}(z)$. Note that the empirical relationship proposed by Jardani et al. (2007) (Eq. (13)) does not account for any salinity-dependence on $\bar{Q}_v^{\text{eff}}$ (see section 2.2 and Fig. 6b). The predicted SP signals (Figure 7) decrease in amplitude with depth and the general shape of the simulated SP signals correspond fairly well to the measurements, but the magnitudes are far too low (1 to 2 orders of magnitude and a corresponding RMSE of 23.7 mV). This result is in agreement with previous field-based studies (e.g., Linde et al., 2011).

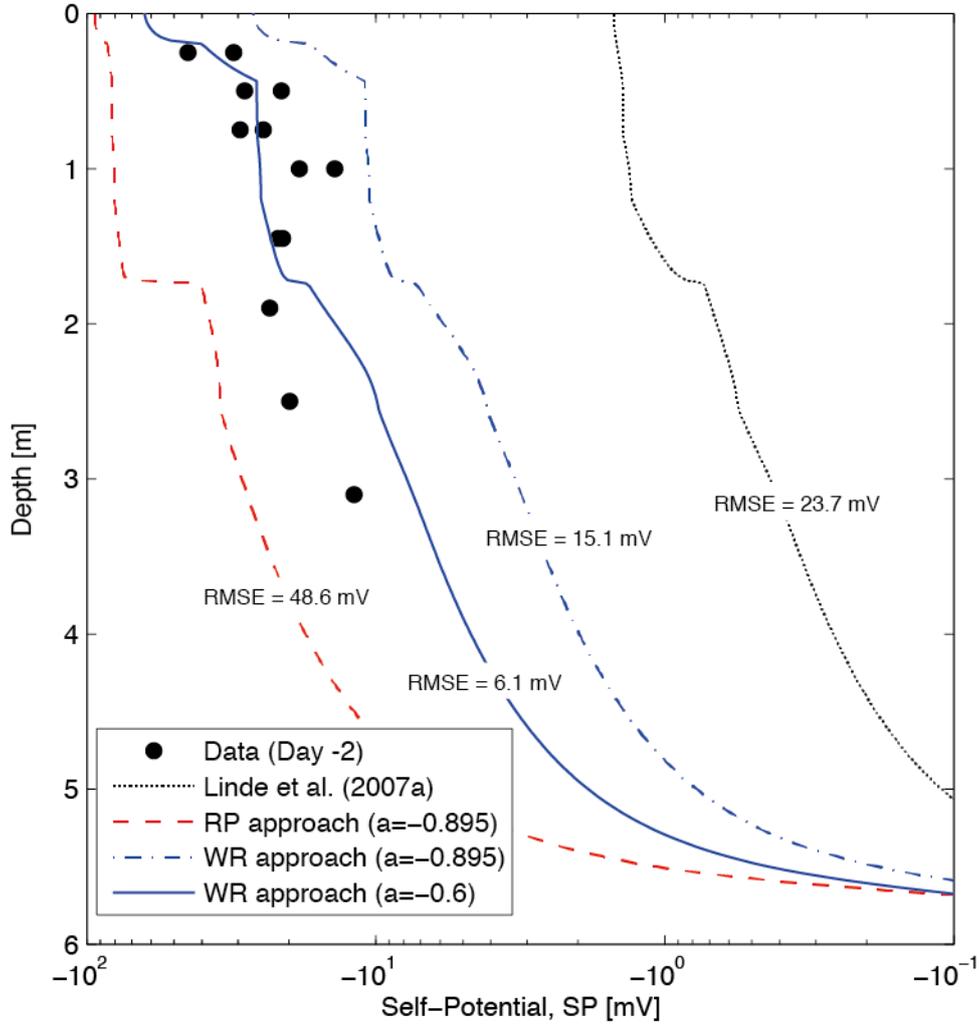

**Figure 7:** Comparison between the measured SP signal before the tracer injection and predictions based on different models describing $\bar{Q}_v^{eff}(S_w)$. The black dashed line is based on the volume-averaging approach of Linde et al. (2007a) using Eqs. (13) and (15) and the predictions are far too low. The red and blue lines correspond to flux-averaging based on the RP approach ($f_D^{RP}$) and the WR approach ($f_D^{WR}$) functions, respectively, scaled with the expected value at saturation. The plain lines and the blue point-dashed line present predictions of the WR approach for $a$ = -0.895 and -0.6 in Eq. (14) to obtain the value at saturation, respectively (see Fig. 6). RMSE between the model simulations and the measurements are indicated on the corresponding curves.

Next, we evaluated the RP and WR approaches by Jougnot et al. (2012). These formulations allow accounting for the salinity dependence on $\bar{Q}_v^{eff}$ (see section 2.2 and Fig. 1). We compare these results with the empirical relationship by Linde et al. (2007b) (Eq. (14) and Fig. 6a) that describes the effect of pore water conductivity at saturation. Based on the RP and WR approaches, Figure 6b provides the predicted response of how the effective excess charge at saturation varies with the pore water conductivity; Jardani et al. (2007) relationship (Eq. (13)) is not salinity-dependent. Figure 6c presents the corresponding variations of the streaming potential coupling coefficients. The streaming potential coupling coefficient predicted by the Jardani et al. (2007) relationship overlap well with Eq. (14) in a narrow range of pore water conductivities (roughly $\sigma_w \in [0.01\,;0.5]$ S m$^{-1}$). The salinity dependence of $C_{EK}$ trend is well captured by the RP and WR approaches for $\sigma_w > 0.03$ S m$^{-1}$ (Fig. 6c). In the following, we scale the calculated relative effective excess charge functions obtained by the RP and WR models with the prediction at saturation based on Eq. (14).

Figure 7 provides a comparison of the SP data with the simulation results based on different $\bar{Q}_v^{\mathrm{eff}}(S_w, C_w)$ relationships. The RP approach overestimates the amplitudes (RMSE = 48.6 mV), while the WR approach underestimates them (RMSE = 15.1 mV). The best agreement with the measured data (RMSE = 6.1 mV) is obtained by scaling $\bar{Q}_v^{\mathrm{eff,rel}}$ obtained by the WR approach using $a = -0.6$ in Eq. (14) as shown in Figure 6a. In the following, we consider only the WR approach scaled with the saturated value from Eq. (14) with $a = -0.6$.

### 4.4. SP responses during the tracer test

The infiltration of the high concentration electrolyte generates Na$^+$ and Cl$^-$ concentration gradients, which results in differential diffusion (see section 2.2) and a non-negligible electro-diffusive contribution ($J_S^{\mathrm{diff}} \neq 0$ A m$^{-2}$ in Eq. (20)).

Figure 8a-d shows the raw SP data at the same four times as in Figure 5 together with the predictions based on the WR approach. The influence of the saline tracer injection is evident in the raw data (Fig 8b, c, and d). The SP signal magnitudes diminish, as predicted by theory ( $\bar{Q}_v^{\text{eff}}$ diminishes and $\sigma_w$ increases with an increasing salinity); this behavior is clearly seen in the SP model predictions. The raw data also display time-varying gradients between 0 and 1.5 m depth that are likely caused by electro-diffusion. Even if the modeled electro-diffusive contributions have the right shape, we find that the magnitudes are far too low. To better highlight the modeled shape, we have in Figure 8e-h enhanced the simulated electro-diffusive contribution with a factor of 7. This enhancement factor is chosen in the following to obtain magnitudes in accordance with the field data and it has no physical meaning. The simulated water flux and ionic concentrations are displayed in Fig. 9a and b, respectively, and the corresponding simulations of the electrokinetic ($SP^{\text{EK}}$) and electro-diffusive ($SP^{\text{diff}}$) contributions are displayed in Fig. 9c and d.

The electrokinetic behavior (Fig. 9c) is clearly different before (Day -2) and after (Days +20, +25, and +40) the tracer injection. The saline tracer increases the pore water salinity, and consequently the electrical conductivity. This induces an amplitude diminution of the potential amplitude (Eq. (20)), which is enhanced by the resulting diminution of $\bar{Q}_v^{\text{eff}}$ (see Figs. 1 and 7), and thus also a reduction of $J_S^{\text{EK}}$ (Eq. (11)). This explains why $SP^{\text{EK}}$ is almost constant where the tracer concentration is very high (e.g., between 0 and 1.2 m depth on day +20).

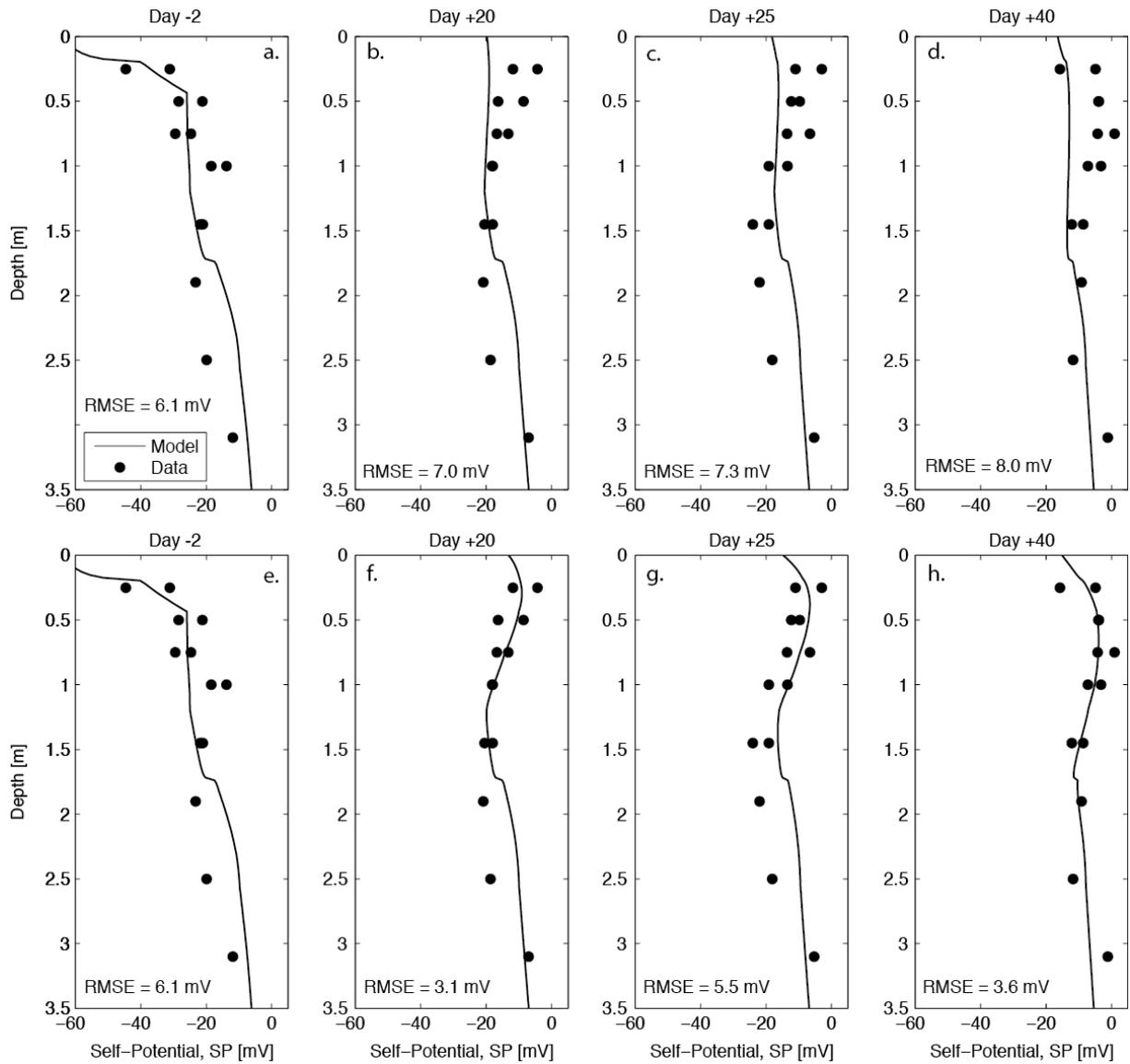

**Figure 8:** Comparison between the measured and simulated SP signal prior (Day -2) and during (Day +20, +25, and +40) the tracer test: the black dots correspond to the measured data and each column to a day. The solid lines in the upper row (a, b, c, and d) correspond to the raw simulated responses (i.e. without enhancement factor), while in the lower row (e, f, g, and h) an enhancement factor of seven for the electro-diffusive contribution was applied. The RMSE between model simulations and measurements are indicated for each subplot.

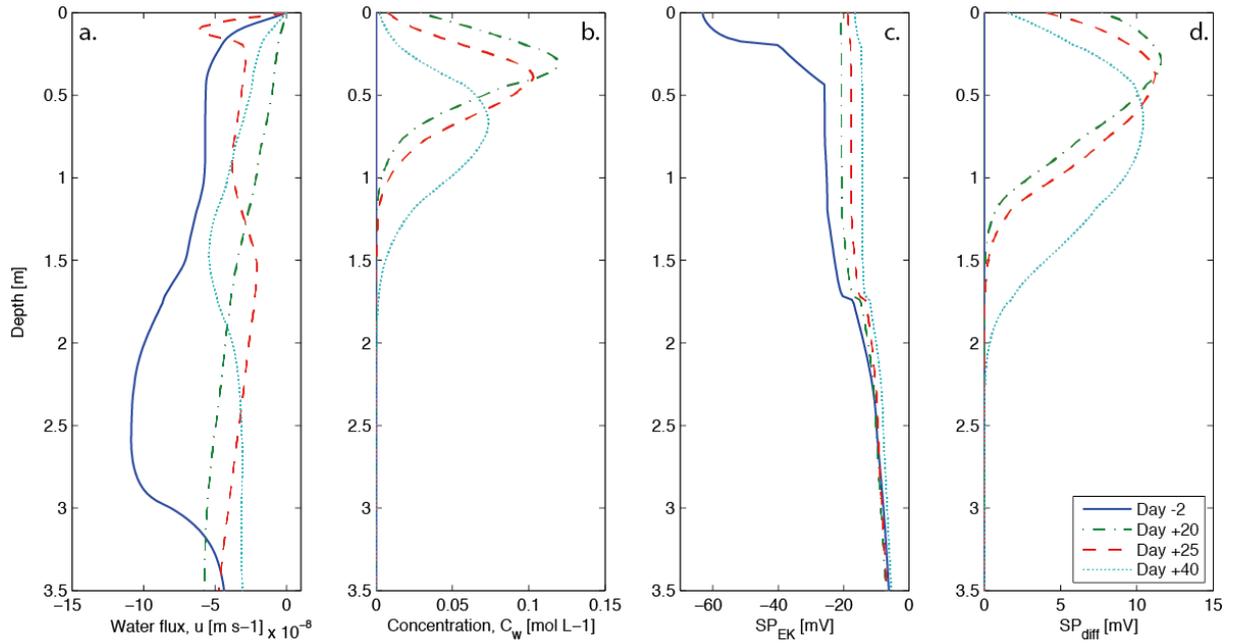

**Figure 9:** Vertical distribution of the (a) water flux, (b) pore water concentration, and the (c) electrokinetic and (d) electro-diffusive contributions to the total SP signal. Note that the enhancement factor of seven is applied to the electro-diffusive contribution (see Fig. 8f, g, and h).

The electro-diffusive contribution to $SP^{diff}$ is calculated from the right term in Eq. (20) by using the simulated NaCl concentration ($C_{Na}$ and $C_{Cl}$ from Hydrus 1D transport solution), the water saturation, and the electrical conductivity. In order to determine the macroscopic Hittorf number, we tested two models: $T_j^H(S_w)$ from Revil and Jougnot (2008) and the simpler $T_j^H(S_w) = t_j^H$ used by Jougnot and Linde (2013) (i.e., the same microscopic Hittorf number as in the pore water). The Revil and Jougnot (2008) model predicts a negative $SP^{diff}$ contribution, which is in contradiction with the experimental data (Fig. 8f, g, and h). This might be due to an over-estimation of the influence of the electrical double layer and its effect on the anion mobility that results in anionic exclusion effects. The microscopic Hittorf number predicts a positive $SP^{diff}$ contribution, but the predicted signals were (as stated above) approximately 7 times smaller than the experimental data (Fig. 8b, c, and d). The underlying reasons for this discrepancy is not well understood at the moment. It can partly be attributed to differences between our electrical resistivity predictions close to the surface (between 0 and 1.5 m depth) and those obtained by the ERT inversions (see Fig. 5). It is in this region that the concentration gradients are the largest and the electro-diffusive contribution is the most important. We can also not exclude that our calculation of the source term in eq. (20) is only partially valid in partially saturated media. This subject has not been addressed in detail in the literature and would be an interesting subject for further studies. We can also not exclude that small-scale heterogeneity, 3-D effects, and preferential flow paths that are not considered in our hydrological model are partly responsible for this discrepancy.

## 5. Discussion

In this work, we compared self-potential field data with modeling results in a partially saturated soil both before and during saline tracer infiltration. Prior to the saline tracer injection, we find that the electrokinetic contribution can explain the raw self-potential SP data quite well. This is the first such field demonstration that has been presented in the literature. As the tracer propagates in the medium, a second source type (electro-diffusion) is needed to explain the measurements. For a downward flux, the $SP^{EK}$ (Fig. 9c) can only be negative, and a positive signal contribution (such as $SP^{diff}$, Fig. 9d) is needed to explain the change in shape of the vertical self-potential profile when the tracer is present in the soil (between 0 and 2 m depth: Figs. 8f, g, and h).

It is clear that the increase of $\bar{Q}_v^{eff}$ with decreasing saturation predicted by Linde et al. (2007a) is insufficient to explain the observed SP amplitudes (Fig. 6). This finding is consistent with recent studies (e.g., Linde et al., 2011; Jougnot et al., 2012). The RP and WR approaches proposed by Jougnot al. (2012) provide significantly better predictions when scaling the relative excess charge function ($\bar{Q}_v^{eff,rel}(S_w)$) with a value at saturation $\bar{Q}_v^{eff,sat}$ in analogy with the relative permeability function (Eq. (4)). This yields simulated results that reproduce the vertical distribution of the electrokinetic contribution. Prior to the tracer injection (Day -2), the SP generation can thus be simulated by only considering the electrokinetic contribution ($SP^{EK}$ from Eq. (11)).

To determine a correct $\bar{Q}_v^{eff,sat}$, we had to account for the influence of salinity upon the excess charge density in the pore water. The widely used approach to determine $\bar{Q}_v^{eff,sat}$ from permeability (Eq. (13)) proposed by Jardani et al. (2007) does not account for the salinity effects illustrated in Fig. 1. Figure 6c shows that the saturated streaming potential coupling coefficients obtained from Eq. (13) overlap well with the empirical relationship proposed by Linde et al. (2007b; Eq. (14)) for typical pore water conductivities ($\sigma_w \in [0.01 ; 0.5]$ S m$^{-1}$). This suggests that this empirical relationship is useful for environmental studies under typical salinities, but is of somewhat limited value at low and high salinities. The salinity dependence of $C_{EK}$ trend is well captured by the RP and WR approaches over a wider salinity range when $\sigma_w > 0.03$ S m$^{-1}$.

The SP data are also strongly influenced by the saline tracer, but the present theory is unable to accurately simulate the resulting magnitudes. When the tracer propagates in the medium, the strong concentration gradients make the electro-diffusive contribution quite significant. It is still an open question how to best model the electro-diffusive contribution under partially saturated conditions. By using a saturation-independent Hittorf number (its value in the pore water), we explain the shape of the SP data after tracer injection (Fig. 8f, g, and h), but the predicted amplitudes appear to be off by a factor 7. This discrepancy is partly attributed to the fact that our predicted electrical conductivities are too high in the tracer-occupied region (Fig. 5b, c, and d). We can also not exclude 3D effects related to the saline tracer plume. Other possibilities relate to the interaction between the salt and the soil matrix or to trapping effects as proposed by Maineult et al. (2006) to explain electro-diffusive signal ($SP^{diff}$) amplitudes larger than those predicted by theory. Furthermore, only few works have focused on the salinity effect on the Hittorf numbers (e.g., Gulamali et al., 2011) and the only model to study the saturation effects were proposed by Revil and Jougnot (2008). This model did not provide results consistent with our data. Jougnot et al. (2012) has clearly highlighted the tremendous influence of saturation on streaming potential magnitudes and we suggest that the situation might be similar for the electro-diffusive contribution. This is an important research area to address in future theoretical and laboratory works.

## 6. Conclusion

At the Voulund agricultural test site, Denmark, we carried out the first ever field-based monitoring of the vertical distribution of SP signals before and during a saline tracer test. We first derived a hydrological model by MCMC inversion to enable hydrological predictions in response to precipitation and tracer injections. We then propose a hydrogeophysical modeling framework that accounts for both water saturation and salinity variations on the simulated SP signals. This is accomplished by using the concept of an effective excess charge, which varies with water saturation and salinity. The most satisfying modeling results were obtained by the so-called WR approach that conceptualizes the porous media as a bundle of capillaries with a distribution that is inferred from the water retention function. Prior to the tracer injection, we find that the electrokinetic (related to water flow) contribution can explain both the signal amplitude and vertical distribution of the raw self-potential data. This was possible without processing or correcting our field SP data. After the tracer injection, it is clear that an electro-diffusive contribution (related to differential diffusion at concentration gradients) is present. The predicted shape of the electro-diffusive contribution is in agreement with the field-data, but the magnitudes are too low. We suggest that this is due to inadequacies in the hydrological model and in our petrophysical relationships. This work confirms many theoretical and laboratory findings in showing that the self-potential data is sensitive to water fluxes and concentration gradients. Our theoretical framework is, in principle, able to predict these contributions and show a satisfactory agreement with field data. Nevertheless, accurate predictions of signal magnitudes are complicated by the many subsurface properties (water content, salinity, porosity, pore size distribution, etc.) that affect the data. Future work will focus on strategies to infer long-term infiltration and groundwater recharge at the Voulund agricultural test site using two-years of high-quality monitoring data.

**Acknowledgement**

The authors thank the Danish hydrological observatory HOBE for the access to the site, technical help (especially Lars Rasmussen), and full access to the data. The authors used a version of Maflot (maflot.com) that was kindly provided by I. Lunati and R. Künze and modified by M. Rosas-Carbajal. The authors kindly thank the associated editor J.A. Huismann, S. Ikard, and two anonymous reviewers for their very constructive comments.